\documentclass[aps,prd,twocolumn,amsmath,amssymb,showpacs]{revtex4}
\usepackage{graphicx,slashed,cancel,bbm,hyperref,framed,xcolor,mathrsfs}
\usepackage[all]{xy}
\usepackage[T1]{fontenc}
\newcommand{\be}{\begin{equation}}
\newcommand{\ee}{\end{equation}}
\newcommand{\bea}{\begin{eqnarray}}
\newcommand{\eea}{\end{eqnarray}}

\newcommand{\sfrac}[2]{{\textstyle\frac{#1}{#2}}}

\newcommand{\D}{\mathrm{d}}
\newcommand{\E}{\mathrm{e}}
\newcommand{\I}{\mathrm{i}}

\newcommand{\Tr}{\mathrm{Tr}}
\newcommand{\Lag}{\mathcal{L}}
\newcommand{\V}{{\mathcal{V}}}

\newcommand{\W}{{\mathcal{W}}}
\newcommand{\T}{{\mathcal{T}}}

\newcommand{\C}{c}

\newcommand{\wlp}{P}
\newcommand{\tach}{\theta}
\newcommand{\lang}{\langle\!\langle}
\newcommand{\rang}{\rangle\!\rangle}

\begin{document}
\title{Worldline holography to all orders and higher spins}
\author{Dennis D.~Dietrich}
\affiliation{Institut f\"ur Theoretische Physik, Goethe-Universit\"at, Frankfurt am Main, Germany}
\affiliation{Arnold Sommerfeld Center, Ludwig-Maximilians-Universit\"at, M\"unchen, Germany}
\begin{abstract}
By using the worldline approach to quantum field theory, we demonstrate {\sl to all orders} that the sources of a quantum field theory over Mink$_4$ naturally form a field theory over AdS$_5$. In particular, this holds for higher-spin sources of a free scalar theory. We work entirely within quantum-field theory and do not select a subset of diagrams. As auxiliary fifth dimension Schwinger's proper time is grouped with the physical four spacetime dimensions into an AdS$_5$ geometry. The four-dimensional sources are extended to five-dimensional fields by a Wilson flow (gradient flow). A variational principle for said flow reproduces the corresponding holographic computation.
\end{abstract}
\pacs{
11.25.Tq 
12.40.Yx 
}
\maketitle

\section{Introduction}

Strong interactions are behind a wealth of phenomena, but frequently overwhelm our computational abilities. Holographic approaches promise analytic insight and have been used intensely in quantum chromodynamics (QCD) \cite{Erlich:2005qh,Karch:2006pv,Polchinski:2000uf}, in extensions of the Standard Model \cite{Hong:2006si,Dietrich:2008ni}, condensed-matter physics \cite{Sachdev:2011wg}, and studies of the Schwinger effect \cite{Sato:2013dwa,Gorsky:2001up,Dietrich:2014ala}. Holography is based on the conjectured AdS/CFT correspondence \cite{'tHooft:1973jz,Maldacena:1997re} and its extensions.  All known examples of this correspondence, however, hold for theories with a set of symmetries that are not realised in nature and have a different particle content. Therefore, deformed bottom-up AdS/QCD descriptions are usually considered, which describe the QCD hadron spectrum surprisingly well \cite{Karch:2006pv,Da Rold:2005zs}. They do, however, lack a derivation from first principles. Consequently, it is very important to understand why and when these models represent an acceptable approximation and which features are robust. For some approaches to these questions see \cite{deTeramond:2008ht,deTeramond:2013it,Cata:2006ak}.

Recently, we demonstrated \cite{Dietrich:2014ala,Dietrich:2013kza,Dietrich:2013vla}, how a quantum field theory over four-dimensional Minkowski space naturally turns into a field theory for its sources over AdS$_5$ in the framework of the worldline formalism \cite{Strassler:1992zr} for quantum field theory, and extends even to the non-relativistic case \cite{Son:2008ye,Balasubramanian:2008dm}. Schwinger's proper time naturally becomes the extra dimension.
Here we show that such an AdS$_5$ formulation arises {\sl to all orders} in the elementary fields---matter and gauge. (Previously, we had selected a subset of diagrams based on the observation that at low energies the contributions with the lowest number of exchanged gauge bosons dominate \cite{Okubo:1963fa,Shifman:1978bx,Dietrich:2012un}.)

The paper is organised as follows: In Section II, we will demonstrate, how a quantum field theory over four-dimensional Minkowski space reorganises into a field theory for its sources over five-dimensional anti-de Sitter space {\sl to all orders} in the elementary fields, with the fifth dimension being Schwinger's proper time of the worldline formalism. In Section III, we analyse the example of the free case and higher-spin sources in detail. Section IV is concerned with dualities between pairs of spacetimes other than Mink$_d$ $\leftrightarrow$ AdS$_{d+1}$. The final section concludes the paper. 

\section{Worldline holography}

\subsection{Volume elements}

To all orders, the correlators of the sources $V$ are described by the generating functional
\begin{align}
Z=\langle\E^w\rangle=\int[\D G]\E^{w-\frac{i}{4e^2}\int\D^4\!X G_{\mu\nu}^2}.
\label{eq:genfun}
\end{align}
In order not to shroud the points important for the present discussion we show the relevant expressions for one scalar flavour of mass $m$ and a vector source $V$, which we incorporate in the `covariant derivative' $\mathbbm{D}=\partial-\I\mathbbm{V}$, together with the gauge field $G$, $\mathbbm{V}=G+V$. (Additional sources and/or flavours can be handled straightforwardly. We comment on corresponding extensions below.) Thus,
\begin{align}
w
=
-\frac{1}{2}\Tr\ln(\mathbbm{D}^2+m^2)
\label{eq:scalar}
\end{align}
instead of the 
$
\frac{1}{2}\Tr\ln(\slashed{\mathbbm{D}}^2+m^2),
$
which we would have for fermionic quarks (on which we are commenting below as well). 
Accordingly, we are also not displaying the path-ordering and traces required for non-Abelian gauge and flavour groups.
In the worldline formalism \cite{Strassler:1992zr} analytically continued to Euclidean spacetime $w$ is expressed as \cite{Dietrich:2014ala,Dietrich:2013kza,Dietrich:2013vla}
\begin{align}
w
=
&\int \D^4x_0\int_{\varepsilon>0}^\infty\frac{\D T}{2\C^2T^3}\,\E^{-m^2T}\,\Lag\equiv\int\D^5x\,\sqrt{|g|}\E^{-m^2T}\,\Lag ,
\nonumber
\\
\Lag
=
&\frac{2\C^2\mathcal{N}}{(4\pi)^2}
\int_\mathrm{P}[\D y]\;\E^{-\int_0^T\D\tau[\frac{\dot y^2}{4}+\I \dot y \cdot\mathbbm V(x_0+y)]} ,
\label{eq:lag}
\end{align}
with the metric parametrisation
\be
\D s^2\overset{g}{=}+\frac{\D T^2}{4T^2}+\frac{\D x_0\cdot\D x_0}{\C T},
\label{eq:patch}
\ee
for the identification of which we collect more pieces of information below.
`$\cdot$' stands for the contraction with the flat four-dimensional metric $\eta_{\mu\nu}$, which after the above analytic continuation from the Minkowski metric with the signature $(-,+,+,+)$ is taken to be plus the Kronecker symbol $\delta_{\mu\nu}$. Then \eqref{eq:patch} is an AdS$_{5,0}$ and before the continuation an AdS$_{4,1}$ line element.
$T$ represents Schwinger's proper time introduced to exponentiate the logarithm in (\ref{eq:scalar}). This step gives rise to a factor of $T^{-1}$ in the volume element. The proper-time regularisation $T\ge\varepsilon>0$ is the analogue of the UV-brane regularisation in holography.  
The subsequent translation of the trace into a path integral contributes another factor of $T^{-2}$, due to one additional integration over four-dimensional momentum space relative to the number of integrations over position space.
The Lagrangian density $\mathcal{L}$ consists of the path integral over all closed paths, $y^\mu(0)=y^\mu(T)$.
The interaction piece takes the form of a Wilson loop $\E^{-\I\oint\D y\cdot V}$, making the local invariance under the transformation $V^\mu\rightarrow\Omega[V^\mu+\I\Omega^\dagger(\partial^\mu\Omega)]\Omega^\dagger$ manifest. Thus, hidden local symmetry \cite{Bando:1984ej} is emergent. The $\D^4x_0$ integral translates the above paths to every position in space, $x^\mu\equiv y^\mu+x_0^\mu$. (The translations are the zero modes of the kinetic operator $\partial_\tau^2$. $\dot y\equiv\partial_\tau y$.)
The appearance of the AdS$_5$ metric reflects the symmetries, SO(4,2), of the conformal group of 3+1-dimensional Minkowski space, which are shared by AdS$_5$. 
Here conformal symmetry is broken by the mass $m$. The factor $\E^{-m^2T}$ can be seen as the tachyon potential $\E^{-\theta^2}$ evaluated on the tachyon profile $\theta=m\sqrt{T}$.

$e^w$ contains all powers of $w$. They read
\begin{align}\nonumber
w^n
=&
\prod_{j=1}^n
\int\D^4x_j\int_{\varepsilon>0}^\infty\frac{\D T_j}{2\C^2T_j^3}\,\E^{-m^2T_j}\,\Lag_j
=\\=&
\int d^5x\sqrt{|g|}\E^{-m^2T}
\int\D^{4(n-1)}\hat\Delta
\nonumber\times\\&\times
\prod_{j=1}^n\int_{\frac{\epsilon}{T}}^{1-\frac{\epsilon}{T}}\frac{dt_j}{2\C^2t_j^3}\,\Lag_j
\times\delta\Big(1-\sum_{k=1}^nt_k\Big).
\nonumber\end{align}
In the second step, we introduced absolute, $x_0$, and dimensionless relative coordinates $\hat\Delta$, such that the Jacobian equals $T^{2(n-1)}/2$. Furthermore, we substituted overall proper time $T$ and proper-time fractions $t_j$ using
\be
1=\int\D T\,\delta\Big(T-\sum_{j=1}^n T_j\Big)\prod_{j=1}^n\Big[\int\D t_j\,\delta\Big(t_j-\frac{T_j}{T}\Big)\Big].\label{eq:choco}
\ee
All but the first $\delta$ serve to carry out the $\D T_j$ integrations, leading to an additional factor of $T^n$. There is a factor of $T^{-1}$ from the first $\delta$, which otherwise constrains the integrations over the proper-time fractions $t_j$. Collecting all the powers of $T$, we find 
$
T^{-3n+2(n-1)+n-1}=T^{-3},
$
which enters into the overall volume element $\sqrt{|g|}$. The integration bounds for the $t_j$ are determined by the lower bound $\epsilon$ for the $\D T_j$ integrations. Accordingly, the $T$ integration starts at $n\epsilon$. 
~Thus, also $w^n$, like $w^1$, takes the the form of a Lagrangian density integrated over AdS$_5$.

\subsection{Contractions}

Expressing $\mathcal{L}_j$ with a mass-dimensionless integration variable $\hat y_j=y_j/\sqrt{T_j}$ yields 
\begin{align}
\label{eq:lagj}
&\Lag_j
=
\frac{2\C^2\hat{\mathcal{N}}}{(4\pi)^2}
\int_\mathrm{P}[\D\hat y_j]
\times\\\nonumber
&\E^{-\int_0^1\D\hat\tau_j[\frac{(\partial_{\hat\tau_j}\hat y_j)^2}{4}+\I\sqrt{t_jT}(\partial_{\hat\tau_j}\hat y_j)\cdot\E^{\sqrt{T}(\sqrt{t_j}\hat y_j+\widehat{x_j-x_0})\cdot \partial_{x_0}}\mathbbm V(x_0)]}.
\end{align}
where $\hat\tau_j=\tau_j/T_j$. ($\widehat{x_j-x_0}$ depends only on $\hat\Delta$ not $x_0$.) Hence, every gradient $\partial_{x_j}$ and every $\mathbbm V$ appears with one power of $\sqrt{T}$. This remains true after integrating out the gauge field $G$,
\begin{align}
\nonumber
&\Big\langle\prod_{j=1}^n\Lag_j\Big\rangle
=
\bigg(\frac{2\C^2\hat{\mathcal{N}}}{(4\pi)^2}\bigg)^n
\prod_{j=1}^n\int_\mathrm{P}[\D\hat y_j]
\nonumber\\&\nonumber
\E^{-\int_0^1\D\hat\tau_j [\frac{(\partial_{\hat\tau_j}\hat y_j)^2}{4}+\I\sqrt{t_jT}(\partial_{\hat\tau_j}\hat y_j)\cdot\E^{\sqrt{T}(\sqrt{t_j}\hat y_j+\widehat{x_j-x_0})\cdot \partial_{x_0}}V(x_0)]}
\\&
\Big\langle
\prod_{l=1}^n\E^{\I\oint\D y_l\cdot G(x_l+y_l)}
\Big\rangle,
\label{eq:aint}
\end{align}
as the average $\langle\dots\rangle$ over the product of Wilson loops (with the integration over $G$ carried out) is $T$ independent, homogeneous of degree 0 in $\{x_j+y_j|j=1\dots n\}$, and independent of $x_0$ (because of translation invariance). Due to the homogeneity of degree 0 and the independence from $x_0$, we can replace $x_j+y_j\rightarrow\hat x_j+\hat y_j~\forall j$ while preserving $x_0$, without altering the value of said average.  
Therefore, the powers of $\sqrt{T}$ remain exclusively and consistently associated with every $\partial_{x_j}$ and $V$. 
Consequently, after carrying out the $[\D\hat y]$ integrations, the result will only contain the combinations $T\eta^{\mu\nu}V_\mu V_\nu$, $T\eta^{\mu\nu}V_\mu \partial_\nu$, and $T\eta^{\mu\nu}\partial_\mu \partial_\nu$, which can be combined into $g^{\mu\nu}V_\mu V_\nu$, $g^{\mu\nu}V_\mu \partial_\nu$, and $g^{\mu\nu}\partial_\mu \partial_\nu$, respectively.
Hence, \eqref{eq:genfun} can be expressed as an action for the sources over AdS$_5$.

\subsection{5d dependence}

One way to continue studying (\ref{eq:aint}) is by expanding it in powers of the source $V$. Every power of the source comes with a translation operator $\E^{\sqrt{T}\hat y\cdot\partial}$. (For scalar sources that would be all there is.) When integrating out the coordinate field $\hat y$, the translation operators are pairwise combined into Gaussian smearing operators \cite{Dietrich:2013kza} $\E^{\wlp\Box}$, where $\wlp\propto T$ is the so-called worldline propagator \cite{Strassler:1992zr}.
All sources are accompanied by smearing operators, which endue them with a $T$ dependence, $V(x)\rightarrow \mathcal{V}(x,\wlp)=\E^{\wlp\Box}V(x)$. $\mathcal{V}(x,\wlp)$ is also the solution of the differential equation defining the Wilson flow (gradient flow) \cite{Luscher:2009eq},
\be
\partial_\wlp \mathcal{V}_\nu=\eta^{\mu\kappa}\mathcal{D}_\mu \mathcal{V}_{\kappa\nu}
\label{eq:wilson}
\ee
where $\mathcal{D}=\partial-\I\mathcal{V}$, when expanded to lowest order in the fields,
\be
\partial_\wlp \mathcal{V}_\perp=\Box \mathcal{V}_\perp,
\label{eq:wilsonlin}
\ee
where $\mathcal{V}_\perp$ denotes the transverse components.

Neither the expansion in the power of sources of (\ref{eq:aint}) nor of (\ref{eq:wilsonlin}), however, is manifestly locally covariant. Alternatively, adopting the Fock-Schwinger gauge $(y_j+x_j-x_0)\cdot V(x_j+y_j)=0$, 
we can express the source $V$ entirely by locally covariant objects, namely the field tensor $V_{\nu\mu}$ and a covariant translation operator \cite{Shifman:1980ui},
\begin{align}\label{eq:shifman}
&V_\mu(x_j+y_j)
=\\=
&\int_0^1\D\eta\:\eta\:\E^{\eta\sqrt{T}(\sqrt{t_j}\hat y_j+\widehat{x_j-x_0})\cdot D(x_0)}[\sqrt{Tt_j}\hat y_j\cdot V_{\bullet\mu}(x_0)], 
\nonumber
\end{align}
where $D=\partial-\I V$. `$\cdot V_{\bullet\mu}$' stands for the contraction of the field tensor on the first index. ($\widehat{x_j-x_0}$ depends only on $\hat\Delta$ not $x_0$.) Using (\ref{eq:shifman}) in (\ref{eq:aint}), we can carry out an expansion in field tensors, which will be accompanied by covariant translation operators. For practical reasons, we would further expand the latter in covariant derivatives. Here, an organising principle is afforded by Taylor expanding the integrand in powers of $T$. This amounts to the so-called inverse-mass expansion \cite{Schmidt:1993rk}. Here the aforementioned induced $T$ dependence of the sources is spread over infinitely many orders of the expansion. The gradients come in gradually, order by order, together with the fields. This amounts to perturbatively approximating the profile $\tilde v$, where $\tilde{\mathcal{V}}(q,T)=\tilde V(q)\tilde v(q,T)$ in momentum space, by a polynomial in $\sqrt{T}q$. Among other things doing so makes the $\D T$ integrand badly convergent at large $T$, especially for small $m$ (whence the name inverse-mass expansion).
In this perturbative context, variational perturbation theory \cite{Variation,Kleinert:2010wb} corresponds to admitting a non-trivial profile for a given order in the sources and a subsequent variation of $Z$ with respect to it.  This is equivalent to working with an a priori arbitrary $\tau$ dependence of the source, $V(x)\rightarrow\mathcal V(x,\tau)$, and varying $Z$ with respect to this source. 
Equating the variation to zero identifies where $Z$ is least sensitive to changes of the profile, but it achieves more, as we shall see below.

Interpreting the absence of a fifth vector component as $\mathcal{V}_T=0$ gauge for a five-vector, the latter also satisfies the five-dimensional Fock-Schwinger gauge condition 
$
(Y_j-X_j)\odot\mathcal V(x_j+y_j,\tau_j)
=
0$,
where $X_j=(x_0-x_j,T)$ and $Y_j=(y_j,\tau_j)$. Then the analogue of (\ref{eq:shifman}) is
\begin{align}
\label{eq:5dtransl}
&\mathcal V_\mu(x_j+y_j,\tau_j)
=\\=
&\int_0^1\D\eta\:\eta\:\E^{\eta(Y_j+X_j)\odot{D}(x_0,T)}[Y_j\odot\mathcal V_{\bullet\mu}(x_0,T)].
\nonumber
\end{align}
Thus, $Z$ depends exclusively on 5d locally covariant objects. By absence of the $\mathcal V_T$ component---which, however, represents a consistent gauge condition in the five-dimensional context---$Z$ is expressed in that gauge. Thus, if we extremise $Z$, which has the form of a gauge-fixed action over AdS$_5$, with respect to $\mathcal V$, we get the same locally invariant result independent of the gauge condition. Hence, additionally, the variationally determined profile leads to a manifestly locally invariant result. 
Moreover, while every four-dimensional index comes with a factor of $\sqrt T$, allowing us to put together the $g^{\mu\nu}$ components, every fifth-dimensional index comes with a factor of $T$, needed for assembling $g^{TT}$. 

~\\

\begin{figure}[t]
\centerline{
\includegraphics[width=0.8\columnwidth]{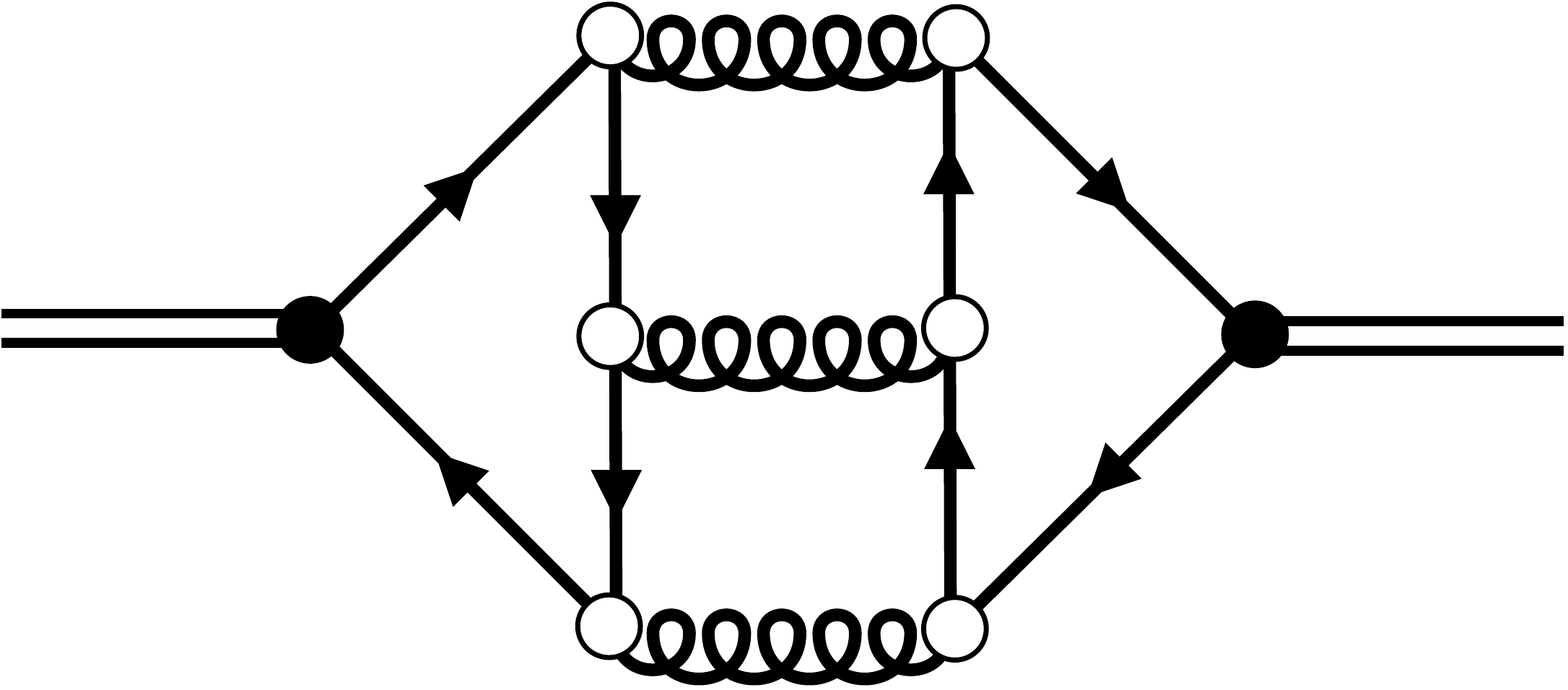}}
\caption{An example for a contribution to the two-point function of unflavoured sources, which is absent for flavoured ones.}
\label{fig:ozi_}
\end{figure}

Diagrams with matter loops with a single outgoing gauge boson that carries colour do not contribute. Moreover, matter loops with sources whose overall net flavour does not vanish, cannot be connected to any other parts by flavour neutral gauge bosons. This causes a difference, for example, between the two-point function of flavoured and unflavoured sources, respectively. The former only have contributions where all sources are connected to one matter loop; for the latter this need not be the case. In the latter case there are diagrams which are connected only by gauge bosons, i.e., which have purely gluonic Fock states (Fig.~\ref{fig:ozi_}), which influences their phenomenology \cite{De Rujula:1975ge}.
Furthermore, matter loops without any attached sources can be included into the average over the gauge Wilson loops, i.e., we do not have to expand the factor $\E^{w_{(0)}}$ in the first place, where the index in parentheses indicates the order in $V$.
Then the expression for the two-point function for a flavoured meson, for example, condenses to
\be\nonumber
\langle w_{(2)} \E^{w_{(0)}}\rangle
=
\int\D^5x\sqrt{|g|}\E^{-m^2T}\langle\mathcal{L}_{(2)}\E^{w_{(0)}}\rangle.
\ee
 [\eqref{eq:choco} has been implemented.] For unflavoured sources, there is a second contribution,
$
\langle w_{(1)}w_{(1)}\E^{w_{(0)}}\rangle.
$

Moreover, that we do not have to expand $\E^{w_{(0)}}$ implies, that with a finite number of external sources, we never have to expand the exponential series to infinite order, and there is no principal issue with convergence. Furthermore, the sourceless matter loops can be excluded from the total proper time $T$ in \eqref{eq:choco}. Thus $\E^{w_{(0)}}$ becomes part of the measure, $\langle\mathcal{O}\E^{w_{(0)}}\rangle=\lang\mathcal{O}\rang$.

In the manifestly locally invariant formalism with non-trivial profile, the lowest-order non-trivial contribution to $Z$ reads \cite{Dietrich:2013kza}
\be
Z_{I\!I}\propto
\int\D^5x\sqrt{|g|}\E^{-m^2T}g^{MN}g^{KJ}\mathcal{V}_{MK}\mathcal{V}_{NJ},
\label{eq:holo}
\ee
in $\mathcal{V}_T=0$.
Capital indices $K,J,M,N,\dots$ count over all five dimensions, while the index `$T$' stands for the fifth component. 
The variation with respect to the source $\mathcal{V}_N$ yields 
\begin{align}
\frac{\partial_T\sqrt{|g|}\E^{-m^2T}g^{TT}g^{\kappa\lambda}\partial_T\V_\lambda}{\sqrt{|g|}\E^{-m^2T}}
+2g^{\mu\nu}g^{\kappa\lambda}\mathcal{D}_\mu\V_{\nu\lambda}
=0
\label{eq:cleom4}
\end{align}
and also
\be
\mathcal{D}\cdot\dot{\mathcal{V}}=0,
\label{eq:cleom}
\ee
if we include a variation with respect to the fifth component.
Thus, the flow $\dot{\mathcal{V}}$ is 4d covariantly conserved like the Wilson flow (\ref{eq:wilson}).
[Also the first addend in \eqref{eq:cleom4} is covariantly conserved.]

For Abelian-like source configurations, $[V_\mu,V_\nu]=0~\forall\mu,\nu$, $Z_{I\!I}$ on a solution of \eqref{eq:cleom4} consists of a surface term,
\be
\breve Z_{I\!I}\propto\int\frac{\D^4q}{(2\pi)^4}q^2|\tilde V_\perp(q)|^2[\ln(q^2\epsilon)+2m^2/q^2+\mathrm{const.}],
\label{eq:saddle}
\ee
where the constant does not depend on $m^2$, $q^2$, or $\epsilon$.
\eqref{eq:saddle} stays finite when $m^2\rightarrow0$. This was also true for the two-point function computed directly, but without maintaining manifest local covariance. To the contrary, $\int\frac{\D T}{T}\E^{-m^2T}$ encountered in the inverse-mass expansion diverges like $\ln m^2$.
This provides another example for the improvement of the infrared convergence through variational perturbation theory \cite{Kleinert:2010wb}. 
These computations for obtaining a manifestly locally invariant profile are equivalent to the corresponding holographic computations \cite{Karch:2006pv,Erlich:2005qh,Dietrich:2008ni}.
The integrand of \eqref{eq:saddle} is proportional to ${q^2|\tilde V_\perp(q)|^2}/{e^2_V}$ of the field theory for the sources, where $e_V$ is the charge of the vector. Thus, by comparison, $e^{-2}_V\propto\ln(q^2/\mu^2)+\ln(\mu^2\epsilon)+\mathrm{const.}$ ($m^2=0$) corresponds to the renormalisation running of the vector charge, with initial conditions given at $\mu^2$. Here, we had absorbed the charge in the source, making that combination renormalisation invariant.

~\\

Also mass renormalisation is captured by the present formalism. In order to see this consider the example of all-orders QED, where the last line of \eqref{eq:aint} is an exponential of double contour integrals $e^2\oint\!\oint\frac{\D y\cdot\D y^\prime}{(y-y^\prime)^2}$ (Feynman gauge). The contributions from where $y\rightarrow y^\prime$ are divergent. (This is the same for any kind of exchanged gauge bosons.) After regularisation, $(y-y^\prime)^2\rightarrow(y-y^\prime)^2+l^2$, the divergent piece goes like $e^2Rl/l^2$, if $y$ and $y^\prime$ are on the same contour with typical size $R$. On the saddle point of the $\D T$ integration the mass and kinetic terms are $\propto Rm$, i.e., linear in $R$ and in $m$. Thus, $Re^2/l$ contributes to mass renormalisation \cite{Affleck:1982,Ritus:1975}. With worldline scale setting, the renormalisation scale is chosen as (proper time)$^{-1/2}$ this leads to a proper-time dependence of $m$. $m\sqrt{T}$ can be seen as the background value of a scalar sourcing $\bar\psi\psi$ (for fermionic matter). For a varying $m=m(T)$, the inverse-mass expansion yields potential and kinetic terms,
\be\nonumber
Z_\theta\propto\int d^5x\sqrt{|g|}\E^{-\tach^2}[1+\#g^{\mu\nu}(\partial_\mu\tach)(\partial_\nu\tach)+\dots],
\ee
where $\tach(T)=m(T)\sqrt{T}$ plays the role of the scalar part of the tachyon in \cite{Bigazzi:2005md}. (We can also include the pseudoscalar part as source for $\bar\psi\gamma^5\psi$.) Mass renormalisation corresponds to a different scaling dimension of the tachyon.

The equations of motion for the mass/tachyon and the charge $e$ (or, alternatively, the dilaton sourcing $G_{\mu\nu}^2$) describe renormalisation effects. The $\beta$ functions are contained within the present formalism \cite{Affleck:1982,Ritus:1975} and do not have to be supplied as additional information. We have motivated mass renormalisation above, but this can be done much more rigorously \cite{Affleck:1982,Ritus:1975}. The renormalisation of the charge is due to loop induced terms $\propto G_{\mu\nu}^2$, analogous to the renormalisation of the vector charge $e_V$.

Using $1/T$ as scale also here, the resulting running gauge coupling $e(T)$ from \eqref{eq:genfun} ends up in the gauge-field averaged Wilson loops in \eqref{eq:aint}. In $e(T)$, $T$ is balanced against a scale set by the initial condition for renormalisation. 
Hence, said average remains homogeneous of degree zero under rescaling of the coordinates. Thus, after integrating out $y$ and $\Delta$ there remains the same function of $e$ as before only that now $e\rightarrow e(T)$. The saddle-point approximation to $e^2\oint\!\oint\frac{\D y\cdot\D y^\prime}{(y-y^\prime)^2}$, for example, always returns $e^2$ times a pure number. Furthermore, the $T$ dependence of $e$ also contributes to the $T$ dependence of the mass/tachyon through $\gamma=\gamma[e(T)]$.

\section{Free case}

Without the gauge interaction the generating functional becomes,
\begin{align}
Z=w
=
&
\int \D^4x_0\int_{\varepsilon>0}^\infty\frac{\D T}{2\C^2T^3}\,\E^{-m^2T}\,\Lag\\\equiv&
\int\D^5x\,\sqrt{|g|}\E^{-m^2T}\,\Lag
 ,
\nonumber
\\
\Lag
=
&\frac{2\C^2\mathcal{N}}{(4\pi)^2}
\int_\mathrm{P}[\D y]\;\E^{-\int_0^T\D\tau[\frac{\dot y^2}{4}+\I \dot y \cdot V(x_0+y)]} .
\label{eq:lag}
\end{align}
It has the form of an action constructed by integrating a Lagrangian density over a five-dimensional space spanned by the four original spacetime coordinates and Schwinger's proper time.
The Lagrangian density consists of a path integral over all closed particle trajectories modulo translations. In this context, the integral over the four-dimensional spacetime corresponds to the translations of the particle trajectories. Translations are the zero modes of the operator $\partial_\tau^2$ in the free worldline action. 

Reexpressing the corresponding Lagrangian density with dimensionless coordinates $\hat\tau=\tau/T$ and $\hat y=y/\sqrt{T}$ yields
\begin{align}
&\Lag
=
\frac{2\C^2\hat{\mathcal{N}}}{(4\pi)^2}
\int_\mathrm{P}[\D\hat y]
\E^{-\int_0^1\D\hat\tau[\frac{(\partial_{\hat\tau}\hat y)^2}{4}+\I\sqrt{T}(\partial_{\hat\tau}\hat y)\cdot\E^{\sqrt{T}\hat y\cdot \partial_{x_0}}V(x_0)]}.
\nonumber
\end{align}
Thus, one sees, that after integrating out the position field $\hat y$ every source $V$ and every gradient $\partial$ is accompanied by a factor of $\sqrt{T}$. Hence, all index contractions are carried out by the combination $T\eta^{\mu\nu}$. Together with the condition $\sqrt{|g|}=(2\C^2T^3)^{-1}$, this selects a metric parametrisation like \eqref{eq:patch}.

The quickest way to compute the Abelian part of the lowest-order term in the inverse-mass expansion is by carrying out an expansion of the worldline action to second order in fields and gradients,
\begin{align}
&\mathcal{N}\int_\mathrm{P}[\D y]\E^{-\int_0^T\D\tau[\frac{\dot y^2}{4}+\I\E^{y\cdot \partial_{x_0}}\dot y\cdot V(x_0)]}
\nonumber\supset\\\supset{}&
-\frac{1}{2}\mathcal N\int_\mathrm{P}[\D y]\int_0^T\D\tau_1\D\tau_2\E^{-\int_0^T\D\tau\frac{\dot y^2}{4}}
\times\\\times{}&
[y_1\cdot \partial_{x_0}\dot y_1\cdot V(x_0)]
[y_2\cdot \partial_{x_0}\dot y_2\cdot V(x_0)]\nonumber
=\\={}&
\frac{1}{2}\int_0^T\D\tau_1\D\tau_2
\label{eq:quick}\times\\&\times\nonumber
(P_{12}\ddot P_{12}\eta^{\mu\kappa}\eta^{\nu\lambda}
+
\dot P_{12}^2\eta^{\mu\lambda}\eta^{\nu\kappa})(\partial_\mu V_\nu)(\partial_\kappa V_\lambda)
=\\={}&
-\sfrac{1}{6}T^2(\eta^{\mu\kappa}\eta^{\nu\lambda}-\eta^{\mu\lambda}\eta^{\nu\kappa})
(\partial_\mu V_\nu)(\partial_\kappa V_\lambda).
\end{align}
Here $y_j=y(\tau_j)$. Already in the first step, we only kept terms in the integrand that give nonzero contributions to the integrals. For the definition of the worldline propagator $P_{12}$ and its derivatives see Appendix \ref{app:wlprop}. In the non-Abelian version, the Abelian field tensor is replaced by the non-Abelian one.

In the expressions for the $N$-point functions the sources always appear in combination with non-trivial profile functions, which correspond to a Wilson-like flow (gradient flow). A general indication for this is, that the ubiquitous translation operators $\E^{y\cdot\partial_{x_0}}$ are integrated into Gaussian smearing operators,
\be
\mathcal{N}\int_\mathrm{P}[\D y]\E^{-\frac{1}{4}\int_0^T\D\tau\dot y^2}\E^{(y_1-y_2)\cdot\partial_{x_0}}=\E^{P_{12}\Box},
\ee
where $P_{12}=O(T)$ and $\Box$ is the flat four-dimensional d'Alembertian. Keeping the full translation operator in the previous computation, for example, would have put exactly this factor inside \eqref{eq:quick}. The inverse-mass expansion sacrifices these profiles for the sake of a manifestly covariant expansion. Above we argued that we can reinstate a profile and keep manifest covariance, effectively by admitting a proper-time dependence of the sources, $V(x_0)\rightarrow\mathcal{V}(x_0,\tau)$, and supplementing it by a variational principle. Thus, we include in the above expansion also all terms up to two derivatives in $\tau$, which yields additionally, 
\begin{align}
&-\frac{1}{2}\mathcal N\int_\mathrm{P}[\D y]\int_0^T\D\tau_1\D\tau_2\E^{-\int_0^T\D\tau\frac{\dot y^2}{4}}
\nonumber\times\\&~\times
(\tau_1-T)\dot y_1\cdot\dot{\mathcal V}_0(\tau_2-T)\dot y_2\cdot\dot{\mathcal V}_0\nonumber
=\\={}&
-\frac{1}{2}\int_0^T\D\tau_1\D\tau_2
(\tau_1-T)(\tau_2-T)\ddot P_{12}
\dot{\mathcal V}_0\cdot\dot{\mathcal V}_0
=\\={}&
-\sfrac{1}{12}T^3\dot{\mathcal V}_0\cdot\dot{\mathcal V}_0.
\end{align}
This is equivalent to replacing 
$y\cdot\partial_\mu F(x)|_{x=x_0}\rightarrow[y_1\cdot\partial_\mu+(\tau_1-T)\partial_\tau]\mathcal{F}(x,\tau)|_{x=x_0}^{\tau=T}$, which is expected from \eqref{eq:5dtransl}.
The above expression is independent of the expansion point, which we here chose as $T$; it only appears in the argument of the source. In case we chose another expansion point, we could change their argument to $T$ by a change of integration variable. Then, after putting all terms together,
\begin{align}
Z_2^{\partial^2}
={}&
-\frac{1}{3}\frac{1}{(4\pi)^2}\int\D^5x\sqrt{|g|}\E^{-m^2T}
\nonumber\times\\&\times
(g^{MK}g^{NJ}-g^{MJ}g^{NK})
(\partial_M\mathcal V_N)(\partial_K\mathcal V_J),
\label{eq:vector}
\end{align}
for $c=8$ all prefactors are absorbed in the metric $g$ and the final expression constitutes the kinetic term for a vector in axial gauge $\mathcal V_T=0$ over AdS$_{5,0}$ before undoing the analytic continuation and AdS$_{4,1}$ afterwards.
Conformal symmetry is broken, as already in the classical theory, by the mass $m$, which here takes the guise of a tachyon potential $\E^{-\theta^2}$ evaluated on the tachyon profile $\theta=m\sqrt{T}$. Additionally, even in the free case the charge of the source is renormalised. [See the discussion after \eqref{eq:saddle}.]

\subsection{Higher spins}
Here we consider a source $W_{\mu_1\dots\mu_L}$ of rank $L$; symmetric in all indices, ${W}_{\mu_1\dots\mu_L}={W}_{(\mu_1\dots\mu_L)}$; traceless, $\eta^{\mu_1\mu_2}W_{\mu_1\dots\mu_L}=0$; and transverse, $\partial_\nu\eta^{\nu\mu_1}W_{\mu_1\dots\mu_L}=0$. The Lagrangian takes the form
\be
\Lag
=
\frac{2\C^2\mathcal{N}}{(4\pi)^2}
\int_\mathrm{P}[\D y]\;\E^{-\int_0^T\D\tau[\frac{\dot y^2}{4}-(-\I \dot y \cdot)^L W(x_0+y)/L!]} ,
\ee
where $(\dot y \cdot)^L W$ denotes the $L$-fold contraction of $W$ with $\dot y$. The contribution to the two-point function comes from
\begin{align}
\Lag_2
=
\frac{(-\I)^{2L}}{2(L!)^2}\frac{2\C^2\mathcal{N}}{(4\pi)^2}
\int_0^T\D\tau_1\D\tau_2\int_\mathrm{P}[\D y]\;\E^{-\int_0^T\D\tau\frac{\dot y^2}{4}}
\nonumber\times\\\times
(\dot y_1\cdot)^LW_1(\dot y_2\cdot)^LW_2.
\end{align}
where $y_1=y(\tau_1)$. Expanding up to the second order in four-gradients yields
\begin{align}
\Lag_2^{\partial^2_x}
={}&
\frac{(-\I)^{2L}}{2(L!)^2}\frac{2\C^2\mathcal{N}}{(4\pi)^2}
\int_0^T\D\tau_1\D\tau_2\int_\mathrm{P}[\D y]\;\E^{-\int_0^T\D\tau\frac{\dot y^2}{4}}
\nonumber\times\\&\times
\{W_0(\cdot\dot y_1)^LW_0(\cdot\dot y_2)^L
+\\&~+\nonumber
[y_1\cdot\partial_{x_0}W_0(\cdot\dot y_1)^L][y_2\cdot\partial_{x_0}W_0(\cdot\dot y_2)^L]
\},
\end{align}
where we dropped the first order in the gradients and terms where both gradients act on the same source source, as they give zero after integration also taking into account the tracelessness of $W$. $W_0=W(x_0)$. Carrying out the $[\D y]$ integration leads to 
\begin{align}
\Lag_2^{\partial^2_x}
={}&
\frac{(-\I)^{2L}}{2L!}\frac{2\C^2}{(4\pi)^2}
\int_0^T\D\tau_1\D\tau_2
[\ddot P_{12}^L W\cdot^LW
\nonumber-\\&-
\ddot P_{12}^LP_{12}\eta^{\mu\nu}(\partial_\mu W)\cdot^L(\partial_\nu W)
\nonumber-\\&-
L\ddot P_{12}^{L-1}\dot P_{12}^2\eta^{\mu\lambda}\eta^{\nu\kappa}(\partial_\mu W_\kappa)\cdot^{L-1}(\partial_\nu W_\lambda)
],
\end{align}
where we have dropped the index 0 to counteract the proliferation of indices. `$\cdot^L$' stands for the $L$-fold contraction of the $2\times L$ suppressed indices with the inverse flat metric $\eta^{\cdot\cdot}$. $P_{12}$ stands for the worldline propagator and $\dot P_{12}$ ($\ddot P_{12}$) for its first (second) derivative with respect to its first argument. (See Appendix \ref{app:wlprop}.) Performing the $\D\tau_j$ integrations leads to
\begin{align}
\Lag_2^{\partial^2_x}
={}&
-\frac{2^{L-1}T^{3-L}}{6L!}\frac{2\C^2}{(4\pi)^2}
\times\\&\times
(\eta^{\mu\nu}\eta^{\kappa\lambda}-L\eta^{\mu\lambda}\eta^{\nu\kappa})
(\partial_\mu W_\kappa)\cdot^{L-1}(\partial_\nu W_\lambda).\nonumber
\end{align}
There are no contributions from $\tau_1=\tau_2$ in the presence of $P_{12}$ or $\dot P_{12}$, since $P_{11}=0=\dot P_{11}$. 
Elsewhere, we regularise powers of coincident $\delta$ distributions according to $[\delta(\tau_1-\tau_2)]^l\rightarrow\delta(\tau_1-\tau_2)/T^{l-1}$, which is tantamount to $\ddot P_{12}^L\rightarrow(-2/T)^{L-1}\ddot P_{12}$.

In the next step, we again admit profiles in the proper-time direction, $W(x_0)\rightarrow\T(x_0,\tau)$, and expand to second order in proper-time gradients,
\begin{align}
\Lag_2^{\partial^2_T}
={}&
\frac{(-\I)^{2L}}{2L!}\frac{2\C^2}{(4\pi)^2}\Big(-\frac{2}{T}\Big)^{L-1}
\dot\T_0\cdot^L\dot\T_0
\nonumber\times\\&\times
\int_0^T\D\tau_1\D\tau_2
(\tau_1-T)(\tau_2-T)\ddot P_{12}
=\\={}&
-\frac{T^{4-L}2^{L-1}}{12L!}\frac{2\C^2}{(4\pi)^2}
\dot\T_0\cdot^L\dot\T_0
\end{align}
where $\T_0=\T(x_0,T)$ and $\dot\T_0=\partial_T\T_0$. Once more, we dropped terms in the integrand that integrate to zero. 
Finally, after collecting all terms, we change the mass dimension of the source to $L$ by rescaling with powers of $T$, $\T=T^{L-1}\W$, such that
\begin{widetext}
\begin{align}
Z_2^{\partial^2}
={}&
-\frac{2^{L-1}}{(4\pi)^2 3L!}\int \D^4x_0\int_{\varepsilon>0}^\infty\frac{\D T}{2T^3}\,\E^{-m^2T}
\{T^{3-L}
(\eta^{\mu\nu}\eta^{\kappa\lambda}-L\eta^{\mu\lambda}\eta^{\nu\kappa})
(\partial_\mu\T_\kappa)\cdot^{L-1}(\partial_\nu\T_\lambda)
+
\sfrac{1}{2}T^{4-L}
\dot{\T}\cdot^L\dot{\T}\}
=\\={}&
-\frac{2^{L-1}}{(4\pi)^2 3L!}\int \D^4x_0\int_{\varepsilon>0}^\infty\frac{\D T}{2T^3}\,\E^{-m^2T}
\times\\&\times
\{T^{L+1}
(\eta^{\mu\nu}\eta^{\kappa\lambda}-L\eta^{\mu\lambda}\eta^{\nu\kappa})
(\partial_\mu\W_\kappa)\cdot^{L-1}(\partial_\nu\W_\lambda)
+
\sfrac{1}{2}T^{L+2}
\dot{\W}\cdot^L\dot{\W}
+
\sfrac{1}{2}T^L(L-1){(1+m^2T)}\W\cdot^L\W
\}
\nonumber=\\={}&\nonumber
-\frac{(2/c)^{L-1}}{(4\pi)^2 3L!}\int \D^5x\sqrt{|g|}\,\E^{-m^2T}
\times\\&\times
\{
(g^{MN}g^{KJ}-Lg^{MJ}g^{NK})
(\partial_M\W_K)(g^{\bullet\bullet})^{L-1}(\partial_N\W_J)
+
4(L-1){(1+m^2T)}\W(g^{\bullet\bullet})^{L}\W
\},
\label{eq:fronsdalpart}
\end{align}
\end{widetext}
where $(g^{\bullet\bullet})^{L}$ stands for the $L$-fold contraction with the five-dimensional inverse metric $g^{MN}$, with which all contractions can be carried out for $c=8$, remembering that all fifth-dimensional components are zero. (We could reexpress the above action using covariant derivatives, which corresponds to shuffling contributions between the kinetic term and the mass term.) 
\eqref{eq:fronsdalpart} is a Fr{\o}nsdal-like action \cite{Fronsdal:1978rb} over AdS$_5$ (after analytic continuation of the time component) for traceless fields without $T$ components and in the presence of a tachyon term. The variation of this action with respect to $W_{\mu_1\dots\mu_L}$ yields the corresponding components of the five-dimensional Fr{\o}nsdal equations once the constraints of tracelessness and transversality have been imposed. There exist several approaches to obtaining the Fr{\o}nsdal equations from an un- or less constrained variational principle, like auxiliary compensator fields \cite{FierzPauli,Singh:1974qz} and the related relaxation to double tracelessness \cite{Fronsdal:1978rb}.

Here, when we evaluated the action on its saddle point in the vector case, we argued, that because the value of the action is independent of the gauge, we could reinstate the $T$ components and perform the computation in another gauge, with the difference, that there is an additional equation from the variation of the $T$ component. Imposing $\mathcal V_T=0$ gauge after the variation it is \eqref{eq:cleom}. In the form $\partial_T(\partial\cdot\mathcal V)=0$ it implies the constancy in $T$ of the divergence of $\mathcal V$. Hence, if $\mathcal V$ is transverse on the boundary, it stays transverse everywhere. 

Analogously, for higher spins, when reinstating the $T$-components in the action formulated with covariant derivatives, the equations with at least one $T$ component read 
\begin{align}
g^{MN}(\nabla_{M}\nabla_{N}\W_{M_1\dots M_L}
-
L\nabla_{M}\nabla_{(M_1}\W_{M_2\dots M_L)N})
=0,\nonumber
\end{align}
where we imposed $\W_T=0$ gauge a posteriori.
Equations with more than two of the $M_j=T$ are identically satisfied by imposing the gauge condition. All the non-zero Christoffel symbols
\be
{\Gamma^T}_{\mu\nu}=\sfrac{4}{2c}\eta_{\mu\nu},~~~
{\Gamma^\mu}_{T\nu}=-\sfrac{1}{2T}\delta^\mu_\nu={\Gamma^\mu}_{\nu T},~~~
{\Gamma^T}_{TT}=-\sfrac{1}{T},
\label{eq:christoffel}
\ee
have an odd number of components in the $T$ direction. The equation with a single $M_j=T$ can thus only have one of them combined with a partial derivative. The equation with two indices in the $T$ direction gets contributions from two Christoffel symbols or two partial derivatives. The latter equation implies
$
\frac{\partial\ln\mathrm{tr}\W}{\partial\ln T}=\mathrm{const.},
$
while in the former a sum over multiples of $\mathrm{div}\,\W$, $\frac{\partial\mathrm{div}\W}{\partial\ln T}$, and $\mathrm{grad}\,\mathrm{tr}\,\W$ vanishes. Therefore, if $\W$ is transverse and traceless on the boundary, it will remain so everywhere. If we had enforced transversality from the very beginning of the derivation of the effective action, the cross term would be absent and the previous equations of motion would simplify to
\begin{align}
0&=
g^{MN}\nabla_M\nabla_N\W_{TTM_3\dots M_L}
=\\&=
g^{\mu\nu}{\Gamma^{\kappa}}_{\mu T}{\Gamma^{\lambda}}_{\nu T}\W_{\kappa\lambda\mu_3\dots\mu_L}
=
\mathrm{tr}\,\W/(4T^2),\nonumber
\end{align}
and
\begin{align}
0&=
g^{MN}\nabla_M\nabla_N\W_{TM_2\dots M_L}
=\\&=
-g^{\mu\nu}(\partial_\mu{\Gamma^\kappa}_{\nu T}+{\Gamma^\kappa}_{\mu T}\partial_\nu)\W_{\kappa\mu_2\dots\mu_L}
=
\mathrm{div}\,\W/T.\nonumber
\end{align}
Thus, in that case the divergence and the trace would be forced to vanish identically, a fact that had been observed before in \cite{deTeramond:2013it}. 

Independent of how we enforce transversality and tracelessness, our worldline holographic equations of motion for the transverse traceless modes are given by
\begin{align}
\Big[-\frac{T^{1-L}}{\E^{-m^2T}}\partial_T\frac{\E^{-m^2T}}{T^{1-L}}\partial_T-\frac{c}{4}\frac{\Box}{T}
+\frac{L-1}{T^2}(1+m^2T)\Big]\W_\perp\nonumber\\=0.
\label{eq:holocleom}
\end{align}
Exploring the behaviour at small values of $T$ with a power-law ansatz, $\W_\perp\propto T^\alpha$ leads to the following characteristic equation
\be
-\alpha(\alpha-2+L)+L-1=0
\ee
with the solutions
\be
\alpha=1~~~\&~~~\alpha=1-L.
\ee
For spinless elementary matter these powers coincide with those found in \cite{deTeramond:2008ht,Karch:2006pv}. The space of solutions for the Fourier transform $\tilde\W(q,T)$ is spanned by $M(L-c\frac{q^2}{4m^2},L+1,m^2 T)T$ (belonging to $\alpha=1$) and $U(L-c\frac{Q^2}{4m^2},L+1,m^2 T)T$ (belonging to $\alpha=1-L$), where $M$ and $U$ are the Kummer functions and $Q^2=-q^2$. The boundary condition for the profile at small values of $T$ is $\lim_{T\rightarrow0}T^{L-1}\tilde\W=\tilde W$, where $\tilde W$ is the Fourier transformed original source. The solutions are polynomial and square normalisable for the discrete values $c\frac{Q^2}{4m^2}-L=n\in\mathbbm{N}_0$, i.e., for $cQ^2=4m^2(n+L)$.

Taking stock, the present framework maps a free scalar quantum field theory on Mink$_4$ with sources of any spin onto a field theory for the sources on AdS$_5$. Such a duality was conjectured to exist for four-dimensional conformal theories \cite{Sundborg:2000wp}. Admitting a mass for the elementary matter breaks conformality and shows up as a tachyon profile in five dimensions. Hence, the value of the parameter $m$ can be interpreted as boundary value for a growing tachyon over an AdS background. From this vantage point, the five-dimensional dual remains one of higher-spin fields over AdS with a certain field taking on a nonzero background value. [In fact, for $L=0$ $\W$ is the tachyon squared $\theta^2$ and  \eqref{eq:holocleom} admits the solution $\theta^2=\W_{L=0}\propto T$ independent of $m^2$. Hence, $m^2$ can be instated as initial value $\lim_{T\rightarrow 0}(\W T^{L-1=-1})=m^2$.] The absorption of the exponential factor in the fields and/or metrics and/or coordinates fails beyond the quadratic level (in the action).

The rank-two sources capture the deviation from the AdS geometry. If the solutions of their equations of motion acquire a nonzero background value in the presence of the tachyon profile, this corresponds to a new effective geometry. Vice versa the scalar profile is influenced by the new effective geometry by interaction terms of the rank-zero and rank-two sources. At the linearised level the equations of motion for the rank-two source possesses the general exclusively $T$-dependent solutions without Lorentz breaking,
\be
\W_{\mu\nu}=\eta_{\mu\nu}\frac{c_1+\frac{c_2}{m^4}[1+(1-m^2T)\E^{+m^2 T}]}{cT}.
\ee
Matching the total metric to the boundary metric sets $c_1=0$. The $c_2$ term in the numerator only contributes from $O(T^2)$ onwards. A nonzero value of $c_2$ corresponds to a modified action. (For values of $c_2$ outside the interval $[-\frac{1}{2};0]$ the effective metric and the corresponding Ricci scalar become singular at some positive $T$. For values inside $[-\frac{1}{2};0[$ the Ricci curvature starts at the constant negative AdS value and grows more and more negative, asymptotically $\propto T^2$.) At this level, however, any value of $c_2$ corresponds to a stationary point of the action. To select a minimal action we have to take into account nonlinear terms and also the backreaction of the tachyon. Moreover, the sources of other ranks may get nonzero background values as well. 

\subsubsection{Interactions}

So far, we had mostly looked at two-point functions and kinetic terms. The present formalism also gives a prescription for determining interactions. [One occasion, where we encountered an interaction was inside the fully non-Abelian kinetic term for a vector \eqref{eq:holo}.] Using the above methods, to the lowest order in the gradients the three-point vertex for sources with even $L$, for example, comes out as
\begin{align}
\Lag_3^{\partial^0}
\propto
\W(g^{\bullet\bullet})^\frac{L}{2}\W(g^{\bullet\bullet})^\frac{L}{2}\W(g^{\bullet\bullet})^\frac{L}{2}\circlearrowleft,
\end{align}
where `$\circlearrowleft$' indicates, that the third source is contracted with the first. For vectors interacting with a rank-$L$ source, on the other hand, we find
\be
\Lag_{\V^L\W}^{\partial^0}
\propto
\V^L(g^{\bullet\bullet})^L\W.
\ee
In both cases imposing tracelessness is instrumental.
Any number of terms can be worked out this way.
While this is expected to be possible consistently over an AdS background \cite{Fradkin:1986qy}, there are impediments over others like dS or Minkowski. Given that our prescription can yield various $d+1$ dimensional spacetimes (see below) a thorough study of interactions in our framework is an important future task.	

\section{Other geometries}

As was just discussed, backreacting the effective metric made up of the original AdS metric together with the contributions from the rank-two sources, can lead to a modified effective geometry. Sources of any rank may participate in characterising the effective spacetime. There are, however, more possibilities, which we are going to describe here.

\subsection{Analytic continuations}

Above, we were concerned worldline holography leading from a quantum field theory over four-dimensional Minkowski space to a field theory for its sources over five-dimensional anti-de Sitter space, more accurately AdS$_{4,1}$. In fact, for the worldline approach we first Wick rotated to Euclidean space and from there found a link to five-dimensional hyperbolic space H$_5$, i.e., AdS$_{5,0}$. Subsequently, continuing back the time direction leads to AdS$_{4,1}$. This corresponds to changing $\epsilon_t$ from $-1$ to $+1$ in the metric
\be
\D s^2\overset{g}{=}\epsilon_T\frac{\D T^2}{4T^2}+\frac{\epsilon_t(\D t)^2+|\D \vec x|^2}{\C T},
\ee
while $\epsilon_T=+1$. By another analytic continuation \cite{Maldacena:2002vr} that takes $\epsilon_T$ from $+1$ to $-1$, we could also reach five-dimensional de Sitter space dS$_5$ from H$_5$ and AdS$_{2,3}$ from AdS$_{4,1}$,
\be
\nonumber
\xymatrix{
\mathrm{dS}_5 \ar@{<->}[r]^{+\epsilon_t-} \ar@{<->}[d]_{\underset{+}{\overset{-}{\epsilon_T}}} & \mathrm{AdS}_{2,3}   \ar@{<->}[d]\ar@{<->}[d]^{\underset{+}{\overset{-}{\epsilon_T}}} \\ 
\mathrm{H}_5  \ar@{<->}[r]_{+\epsilon_t-}              & \mathrm{AdS}_{4,1}
}
\ee
These would still be holographic pictures of four-dimensional flat space, but with an analytically continued proper time. (In \cite{Maldacena:2002vr} this is actually done for a radial coordinate $z$, where $z^2\propto T$, which is rotated from real to imaginary. Accordingly, here this corresponds to a change of the sign of $T$ and then appears like doing holography on the other side of the boundary.)

\subsection{Finite temperature}

In thermal field theory in the imaginary time formalism the time direction is compactified with the period given by the inverse temperature. If we run our present machinery straightforwardly we find thermal AdS space, i.e., AdS space with a compactified temporal direction. It is known, however, that there is a second space with the same boundary topology, the AdS black hole \cite{Hawking:1982dh}. In the present formalism the difference between the two spaces does not arise by holographically extending some finite difference present already on the boundary, but on sources of the stress-energy tensor  that vanish on the boundary together with their first derivatives \cite{Bautier:2000mz} and which combine with the `naive' background to yield the five-dimensional metric. Taking these sources into account, both geometries, the thermal AdS with the extra components identically zero, and the AdS black holes, with finite extra components, are stationary points of the action. The relative value of the action then decides about the preferred configuration. It turns out that in the present setting the relative importance of bosonic and fermionic degrees is decisive for which of the two five-dimensional spacetimes is preferred \cite{Dietrich:2015dka}.

\subsection{Non-relativistic case}

The Schr\"odinger equation's conformal Galilean symmetry in three spatial dimensions can be constructed by constraining (\ref{eq:lag}) without the factor $\E^{-m^2 T}$ in 4+1 dimensional Minkowski space by imposing on the light-cone momentum $p^+=m$ \cite{Son:2008ye}. Correspondingly, $x^+$ takes the role of time. This leads to a six-dimensional volume element (two extra dimensions!) $\propto T^{-7/2}$. 
This  is the volume element for the correct metric \cite{Son:2008ye,Balasubramanian:2008dm},
\be
\D^2s\overset{\mathrm{g}}{=}
-\frac{\D T^2}{4T^2}
+\frac{2(\D x^+)^2}{T^2}
+\frac{2\D x^+\D x^--\D\mathbf{x}\cdot\D\mathbf{x}}{T} ,
\ee
where $x^\pm=\frac{x^0\pm x^4}{\sqrt{2}}$ and $x^4$ is the coordinate in the direction of the initially present extra dimension. (The power of $T$ in the denominator of the $(\D x^+)^2$ term can be different \cite{Balasubramanian:2008dm} without influencing the volume element.) In the relativistic case, on top of the factor of $\frac{1}{T}$ from the integral representation of the logarithm, the volume element was due to the mismatch in numbers of position and momentum integrals in the path integral, 
\be
\frac{1}{T}\int\D^4p\,\E^{-Tp^2}\propto T^{-3}\propto\sqrt{|g|}.
\ee
For the present non-relativistic case, we have an analogous mismatch,
\be
\frac{1}{T}\int\D^3\mathbf{p}\,\D p^-\,\E^{-T(2mp^-+\mathbf{p}^2)}\propto T^{-7/2}\propto\sqrt{\mathrm{|g|}}.
\ee

\subsection{General curved spacetime}

If we start from a general curved spacetime,
we have to analyse a spacetime dependent metric $h=h(x)$ in the worldline action. We can still split off a translational coordinate $x_0$, where $x=x_0+y$ and $\dot x_0\equiv0$, such that
\begin{align}
\nonumber
Z_h\propto{}
&\int\frac{\D T}{2T^3}\mathcal{N}\int_\mathrm{P}[\D x]\E^{-\int_0^T\D\tau[\sfrac{1}{4}\dot x\cdot h(x)\cdot\dot x+\I\dot x\cdot h(x)\cdot V(x)]}
=\\={}&\nonumber
\int\frac{\D T}{2T^3}\D^4x_0\sqrt{|h(x_0)|}\,\mathcal N\int_\mathrm{P}[\D y]
\times\\&\times
\E^{-\int_0^T\D\tau[\sfrac{1}{4} \dot y\cdot h(x_0+y)\cdot\dot y+\I\dot x\cdot h(x)\cdot V(x)]},
\end{align}
where we also included a rank-1 source as example.
$x_0$ will generally not be zero modes of the worldline kinetic operator anymore, as $h$ is not constant. The volume element of a Fefferman-Graham metric
\be
\D s^2\overset{H}{=}\frac{\D T^2}{4T^2}+\frac{\D x\cdot h\cdot\D x}{\C T}
\label{eq:H}
\ee
still emerges consistently, since
\be
\sqrt{|H|}=\frac{\sqrt{|h|}}{2c^2T^3}.
\ee
Using Riemann normal coordinates centred at $x_0$, we get to leading order
\begin{align}
Z_h^{(0)}\propto{}
&\int\frac{\D T}{2T^3}\mathcal{N}\int_\mathrm{P}[\D y]_0\E^{-\int_0^T\D\tau[\sfrac{1}{4}\dot x\cdot h(x_0)\cdot\dot x+\I\dot x\cdot h(x_0)\cdot V(x)]}.\nonumber
\end{align}
From thereon the computations proceed as above, since $h(x_0)$ is a constant from the viewpoint of the path integration. We get the previous results with all contractions carried out with the metric \eqref{eq:H}. (In order to see this, it is helpful to work in coordinates that all have the dimension of a length.) At this point, there will be no couplings between the sources and curvature tensors. These will be generated at higher orders of the expansion. Then, however, one must take care of some technical details: On the one hand, the path-integral measure is non-trivial on a curved space, which can be handled by the introduction of additional ghost fields \cite{Lee:1962vm} or alleviated by adopting unimodular gauge, $\sqrt{|h|}=$const. Moreover, depending on how the path integral is carried out in practice, there will be correction terms in the worldline action that depend on the metric $h$ \cite{Mizrahi:1975pw}. In spaces of constant Ricci curvature $R$, they give rise to an overall factor of $\E^{-RT/16}$, which thus resembles a mass term and constitutes part of the source-curvature interactions. In more general spacetimes already at the leading order in an expansion in Riemann normal coordinates this factor would be four-spacetime dependent, $\E^{-R(x_0)T/16}$. 

\section{Short summary}

The purpose of the present paper was to identify such aspects of holography that can be obtained without input from string theory \footnote{While there was no input from string theory to the present derivations, stringy aspects can nevertheless be gleaned. After all, at the latest since \cite{'tHooft:1973jz} we know that at least part of the gauge field averaged Wilson loops, $\langle\dots\rangle$, should have a description in terms of a string theory.}.
Thus, completely within quantum field theory we have demonstrated that a quantum gauge field theory over Mink$_4$ corresponds to a field theory for its sources over AdS$_5$ to {\sl all orders} in the elementary fields, matter and gauge. The fifth coordinate is Schwinger's proper time in the worldline formalism \footnote{The use of the worldline formalism already permitted to derive the Bern-Kosower formalism \cite{Bern:1991aq} without recourse to string theory \cite{Strassler:1992zr,Strassler:1993km}.}
\footnote{The use of the worldline formalism also allows us to make a link \cite{Dietrich:2007vw} to the Gutzwiller trace formula \cite{Gutzwiller:1971fy}, which describes quantum systems through classical attributes, as do holographic computations \eqref{eq:saddle}.}.
The sources receive an induced dependence on this variable corresponding to a Wilson flow (gradient flow). Extremising the action with respect to this flow---which is motivated by demanding a locally invariant result with good infrared behaviour---reproduces the holographic computations.

When starting, for example, from a free scalar theory over Mink$_4$ sourced with tensors of any spin, we obtain a field theory for these higher-spin sources over AdS$_5$. The mass of the elementary matter takes the guise of a tachyon profile, which breaks conformal symmetry in five-dimensions, as already did the elementary mass in four. Additional contributions arise from charge renormalisation.
In an interacting theory, renormalisation of the gauge charge and the elementary mass are further sources of conformal symmetry breaking.

The program generalises straightforwardly to other pairs of spacetimes with $d$ and $d+1$ dimensions, respectively. By various combinations of analytical continuations we can also link Mink$_d$ and Eucl$_d$ to AdS$_{d,1}$, AdS$_{d+1,0}$, AdS$_{2,d-1}$, and dS$_{d+1}$.
At finite temperature the five-dimensional spacetime turns out to be either thermal AdS space or the AdS black hole \cite{Dietrich:2015dka}.
Commencing from a general curved $d$ dimensional manifold, to leading order in Riemann normal coordinates the $d+1$-dimensional metric is the Fefferman-Graham metric for the given $d$-dimensional metric. The subsequent orders lead to source-curvature interactions.

\section*{Acknowledgments}

D.D.D.~would like to thank
Stan Brodsky,
Guy de T\'eramond,
Luigi Del Debbio,
Gerald Dunne,
Gia Dvali,
Daniel Flassig,
Christiano Germani,
C\'esar G\'omez,
Alexander Gu{\ss}mann,
Stefan Hofmann,
Paul Hoyer,
Adrian K\"onigstein,
Sebastian Konopka,
Michael Kopp,
Matti J\"arvinen,
Florian Niedermann,
Yaron Oz,
Joachim Reinhardt,
Tehseen Rug,
Ivo Sachs,
Debajyoti Sarkar,
Andreas Sch\"afer,
Robert Schneider,
Karolina Socha,
Stefan Theisen,
Christian Weiss,
Nico Wintergerst,
and
Roman Zwicky
for discussions. 
The work of the author was supported by the Humboldt foundation and the European Research Council.

\appendix

\section{Worldline propagators\label{app:wlprop}}

We frequently have to compute expectation values of products of position operators $y$ and its first proper-time derivative $\dot y$ over all closed paths with the weight $\E^{-\frac{1}{4}\int_0^T\D\tau\dot y^2}$. It is practical to introduce auxiliary sources $c$ and $s$ in the path-integral and replace $y$ and $\dot y$ with functional derivatives, $y\rightarrow\delta_c$ and $\dot y\rightarrow\delta_s$. The sources will be put to zero after all functional derivatives have been carried out. Then we can  perform the path integral and find
\begin{align}
&\mathcal{N}\int[\D y]\E^{\frac{1}{4}\int_0^T\D\tau (y\cdot\partial_\tau^2y+4c\cdot y+4s\cdot\dot y)}
=\\={}&
\mathcal{N}\int[\D y]\E^{\frac{1}{4}\int_0^T\D\tau (y\cdot\partial_\tau^2y+4c\cdot y-4\dot s\cdot y)}
=\\={}&
\mathcal{N}\int[\D y]\E^{\frac{1}{4}\int_0^T\D\tau[y+2(c-\dot s)\partial_\tau^{-2}]\cdot\partial_\tau^2[y+2\partial_\tau^{-2}(c-\dot s)]}
\times\\&\times
\E^{-\int_0^T\D\tau(c-\dot s)\cdot\partial_\tau^{-2}(c-\dot s)}
=\\={}&
\E^{-\frac{1}{2}\int_0^T\D\tau_a\D\tau_b(c_a-\dot s_a)\cdot P_{ab}(c_b-\dot s_b)}
=\\={}&
\E^{\frac{1}{2}\int_0^T\D\tau_a\D\tau_b(-c_a\cdot c_bP_{ab}-c_a\cdot s_b\dot P_{ab}+s_a\cdot c_b\dot P_{ab}+s_a\cdot s_b\ddot P_{ab})}.
\label{eq:auxgenfunct}
\end{align}
Here $P_{12}=P(\tau_1,\tau_2)$ stands for the worldline propagator and $\dot P_{12}$ ($\ddot P_{12}$) for the first (second) derivative with respect to its first argument. We shall be working in the `centre-of-mass' conventions, where $x_0$ is defined such that $\int_0^T\D\tau y=0$. Then the worldline propagator is translationally invariant in Schwinger's proper time, $P(\tau_1,\tau_2)=P(\tau_1-\tau_2)$ and obeys the equation of motion
\be
\ddot P_{12}=2\delta(\tau_1-\tau_2)-2/T.
\ee
The solution and its first derivative read
\begin{align}
P_{12}&=|\tau_1-\tau_2|-(\tau_1-\tau_2)^2/T,
\\
\dot P_{12}&=\mathrm{sign}(\tau_1-\tau_2)-2(\tau_1-\tau_2)/T.
\end{align}

\end{document}